# Maximizing the Value of Solar System Data through Planetary Spatial Data Infrastructures

Primary Author: **Jani Radebaugh**, Chair, Mapping and Planetary Spatial Infrastructure Team (MAPSIT), Brigham Young University, janirad@byu.edu, (801) 422-9127

COAUTHORS

*The MAPSIT Steering Committee:* **Brad Thomson**, Vice-Chair, University of Tennessee Knoxville; **Brent Archinal**, U. S. Geological Survey; **Ross Beyer**, SETI Institute, NASA Ames; **Dani DellaGiustina**, U of Arizona Lunar and Planetary Laboratory; **Caleb Fassett**, Marshall Space Flight Center; **Lisa Gaddis**, U. S. Geological Survey; **Sander Goossens**, Goddard Space Flight Center; **Trent Hare**, U. S. Geological Survey; **Jay Laura**, U. S. Geological Survey; **Pete Mouginis-Mark**, SOEST, University of Hawaii; **Andrea Naß**, Deutsches Zentrum für Luft- und Raumfahrt; **Alex Patthoff**, Planetary Science Institute; **Julie Stopar**, Lunar and Planetary Institute; **Sarah Sutton**, U of Arizona Lunar and Planetary Laboratory; **David Williams**, SESE, Arizona State University; **Justin Hagerty** (ex officio), Director, Astrogeology Science Center, U. S. Geological Survey; **Louise Prockter** (ex officio), Planetary Data System lead scientist, Lunar and Planetary Institute

ENDORSERS

**Akos Kereszturi**, Research Centre for Astronomy and Earth Sciences; **Anezina Solomonidou**, European Space Agency (ESA/ESAC); **Sebastian Walter**, Freie Universitaet Berlin; **Alessandro Frigeri**, Istituto di Astrofisica e Planetologia Spaziali, INAF, Rome, Italy; **Maria Gritsevich**, 1. Finnish Geospatial Research Institute FGI, Geodeetinrinne 2, 02430 Masala, Finland; 2. Institute of Physics and Technology, Ural Federal University, 620002 Ekaterinburg, Russia; **Jeffrey R. Johnson**, Johns Hopkins University Applied Physics Laboratory; **Jennifer E. C. Scully**, Jet Propulsion Laboratory, California Institute of Technology; **Eleni Ravanis**, European Space Agency (ESA/ESAC); **Ari Koeppel**, Northern Arizona University; **Abigail R. Azari**, Space Sciences Laboratory, University of California, Berkeley; **Audrie M. Fennema**, Lunar and Planetary Laboratory, University of Arizona; **Devon M. Burr**, Northern Arizona University; **Frédéric Shmidt**, GEOPS, Université Paris-Saclay, France ; **Tilmann Denk**, German Aerospace Center (DLR), Germany; **Andrew D. Horchler**, Astrobotic; **Tom Farr**, Jet Propulsion Laboratory; **Emily J. Forsberg**, University of Idaho; **Geoffrey C. Collins**, Wheaton College; **Michael Bland**, Astrogeology Science Center, U. S. Geological Survey; **Kelsi N. Singer**, Southwest Research Institute; **Conor A. Nixon**, NASA Goddard Space Flight Center; **Emily Law**, Jet Propulsion Laboratory, California Institute of Technology; **Brian Day**, NASA Solar System Exploration Research Virtual Institute; **Andrew M. Annex**, Johns Hopkins University; **Devanshu Jha**, College of Engineering, University of Bangalore; **Tim Glotch**, Stony Brook University; **Ken Herkenhoff**, U.S. Geological Survey; **Colin Dundas**, U.S. Geological Survey

*The SBAG Steering Committee:* **Bonnie Buratti**, Chair, Jet Propulsion Laboratory, California Institute of Technology; **Dan Adamo**, Human Exploration Lead, NASA HQ; **Elena Adams**, Technology Lead, Johns Hopkins University Applied Physics Laboratory; **Maitrayee Bose**, *Arizona State University;* **Michael Disanti**, *NASA*, Jessie Dotson, Planetary Defense Lead, *NASA Ames;* **Carolyn Ernst**, Johns Hopkins University Applied Physics Laboratory; **R. Terik Daly**, Johns Hopkins University Applied Physics Laboratory; **David Gerdes**, *University of Michigan;* **Andy Rivkin**, Johns Hopkins University Applied Physics Laboratory; **Jennifer Scully**, Jet Propulsion Laboratory, California Institute of Technology; **Tim Swindle**, Past Chair, Lunar and Planetary Laboratory, *University of Arizona*; **Patrick Taylor**, *Lunar and Planetary Institute*;

# Summary

Planetary spatial data returned by spacecraft, including images and higher-order products such as mosaics, controlled basemaps, and digital elevation models (DEMs), are of critical importance to NASA, its commercial partners and other space agencies (e.g., NASA Data Plan 2014). Planetary spatial data are an essential component of basic scientific research and sustained planetary exploration and operations. The Planetary Data System (PDS) is performing the essential job of archiving and serving these data, mostly in raw or calibrated form, with less support for higher-order, more ready-to-use products. However, many planetary spatial data remain not readily accessible to and/or usable by the general science user because particular skills and tools are necessary to process and interpret them from the raw initial state. In addition, many spatial data do not have uniform spatial scales and/or have not been accurately registered to a standard cartographic base, prohibiting simple tasks such as locating features or measuring distances or elevations. ***There is a critical need for planetary spatial data to be more accessible and usable to researchers and stakeholders.*** A Planetary Spatial Data Infrastructure (PSDI) is a collection of data, tools, standards, policies, and the people that use and engage with them. A PSDI comprises an overarching support system for planetary spatial data. PSDIs (1) establish effective plans for data acquisition; (2) create and make available higher-order products; and (3) consider long-term planning for correct data acquisition, processing and serving (including funding). PSDIs improve the potential for correlating existing and future datasets in ways that will increase the overall scientific return, enable new research activities and inspire the general public and next generation scientists. **We recommend that Planetary Spatial Data Infrastructures be created for all bodies and key regions in the Solar System.** NASA, with guidance from the planetary science community, should follow established data format standards to build foundational and framework products and use those to build and apply PDSIs to all bodies. *Establishment of PSDIs is critical in the coming decade for several locations under active or imminent exploration, and for all others for future planning and current scientific analysis.* Only through these efforts can Solar System spatial data, of great scientific value and acquired at great cost, be used to its maximum potential and help NASA achieve its goals.

# Planetary Spatial Data Infrastructures

**What and Where are Planetary Spatial Data?** Planetary spatial data include those from orbital remote sensing instruments operating at a variety of wavelengths, surface images collected by rovers and landers, and samples collected from known locations on planetary bodies (Laura, et al. 2017; 2018a,b). Planetary spatial data are typically collected in raw form, processed by instrument teams, and ultimately made publicly available via the Planetary Data System (PDS) archives. NASA's PDS manages close to 2 petabytes of data (PDS Roadmap 2017) and large amounts of new data are archived each year by active space missions. In most cases, data stored within the PDS are not ***spatially enabled*** for immediate use by non-expert research scientists. This means that they are not intuitively usable: for instance, uncontrolled data are not accurately registered and may not align correctly for comparing values in different images or data sets. Instead, the burden is on the user to create spatially enabled products. Using the provided metadata, the user must use navigation, spacecraft and instrument positional information, sensor models, radiometric and cartographic processing pipelines, and photogrammetric and radargrammetric control software to create spatially enabled products. Significant expertise is required to perform these operations and interpret the spatial correctness of the products. Some datasets are stored outside of the PDS (e.g., on institutional websites or other repositories). These data may be in formats that are not compatible with the PDS and are of varying quality and usability. These data also often require the user to derive correct and usable products.

**Current Usability of Planetary Spatial Data:** There are numerous efforts within the NASA planetary science community that each focus on some aspect of making data readily usable to the community. Most missions have their own web servers to store raw and processed data and make them available to mission team members. The higher-order products made by these missions may or may not include data that have been accurately registered onto the foundational coordinate reference frame of the target planetary body, which would enable correct location and measurement of features. Instead, mission products are often more stopgap products, intended for planning and first-level science interpretations, and mission lifetimes and funding can end before more extensive image processing and registration can be performed. The PDS has a NASA charter to archive and deliver mission data (e.g., Gaddis et al. 2018) and to focus on long-term preservation (Laura et al. 2018a). However, there is no entity tasked with the laborious and complex process of registering spatial data to the body. These activities are typically left to be funded through PI-level opportunities to generate, calibrate, and rectify planetary spatial data (e.g., such as NASA ROSES PDART), which can lag years behind data acquisition. The typical planetary data user (who often is not an expert in spatial data manipulation) may not be able to process and use raw, archived planetary data without a significant investment of time and funding. While in recent years NASA has been funding the restoration and development of higher-order products that are properly registered and ready-to-use, and these are being archived in the PDS, such data products are rare for many planetary bodies, or were created with outdated standards, coordinate frames, and techniques, or they lack information on accuracy. ***A more concerted, sustainable and focused effort is needed to fill the gaps in usable data and enable more coordinated use of planetary data to address high-priority goals for science and exploration.***

**Planetary Spatial Data Infrastructures:** A Planetary Spatial Data Infrastructure (PSDI) is a series of agreements on standards, institutional arrangements, and policies for describing the bounds within which spatial data planning should occur and for organizing the resulting data in a standardized way so that data are discoverable, accessible and usable in support of a focused endeavor (Laura et al. 2017; Hargitai et al. 2019). The ideal is to create focused PSDIs centered around a body or specific exploration goals or scientific task or research theme. An ideal PSDI should serve a broad community whose members do not need to be experts in spatial concepts and who may not understand the details of storing, finding, and using spatially enabled data (Laura et al. 2017). Most users of planetary spatial data want the data to "just work," and a PSDI should enable this. **Foundational Data** provide the basic positional structure upon which all other data are registered in a PSDI (Duxbury et al. 2002; Archinal et al. 2018). Foundational products include mosaics and other products based on data that are accurately scaled and registered to features on a given planetary surface with known locations. These products, along with their quantitative assessment of spatial efficacy, serve as a base to which additional products can be controlled. Upon this structure sit higher-order, more specialized **Framework Data**. These can be diverse and include multi-temporal remote sensing images, hyperspectral compositional data, nomenclature and map units (Laura et al. 2017). Assembling all accurately registered, foundational data into a structure with an interface that supports straightforward access to products and provides visualization and analysis tools for any user is an important part of a Planetary Spatial Data Infrastructure (PSDI). This philosophy encourages the development of PSDIs for many different Solar System bodies and efforts (Archinal et al. 2017; Laura et al. 2017; Laura et al. 2018a,b; Hargitai et al. 2019). PSDI-related efforts may include gathering all data needed for selection or characterization of a safe landing site or establishing a geodetic coordinate reference frame to support spatially accurate measurement, navigation and traditional mapping techniques. Fusion of datasets allows for co-analysis that greatly enhances the science return. These efforts would support the planning and operation of missions by NASA and other agencies.

## Needs and Proposed Actions

A PSDI would describe the steps needed to correctly obtain and process planetary spatial data, to develop new foundational products, and to create additional usable products necessary for future science and exploration. Currently envisioned, planetary body-focused PSDIs would address the needs of upcoming NASA missions (Archinal et al. 2017; Laura et al. 2018b) by laying out what products exist, which are still needed, and what formats and product types are most usable for the mission and research community. Ideally, PSDIs can contribute to new research discoveries as well as help reveal to NASA where strategic knowledge gaps exist. This knowledge can be used to help decide where it is most scientifically interesting and technically feasible to land, image, or fly past planetary bodies for science or exploration. These efforts must be executed for several locations this decade:

A Europa PSDI would enable critical planning and execution of the upcoming Europa Clipper mission (Laura et al. 2018b); a Mars Jezero crater PSDI would enable effective operation of the Mars 2020 rover, and lunar PSDIs must be created for critical human and robotic landing site operations, all imminent. ***PSDIs enable the best and most effective use of spatial data and are needed for all planetary bodies, at different scales, to support a wide variety of purposes.*** To proceed with creating PSDIs and enhance the science return of NASA spatial data, we make the following recommendations from the Mapping and Planetary Spatial Data Infrastructure (MAPSIT) Roadmap: https://www.lpi.usra.edu/mapsit/roadmap/

**Recommendation I:** NASA missions should be encouraged to obtain high-quality data that can be controlled, registered and incorporated into existing foundational data products, or create foundational data products for new territory, and thus maximize the value of the NASA science return. These activities should be executed at an early level of mission planning. NASA and the community should work with missions to:

- Encourage obtaining data of the highest possible spatial quality, with awareness of current ephemerides and use of spacecraft tracking, so that data can be placed spatially on the body and within the region as accurately as possible.
- Ensure that adequate data calibration plans are in place and executed as needed prior to launch and in flight (NASA could consider providing the calibration service), and that calibration data and documentation are delivered to the public in a timely fashion during the mission.
- Collect data that are synergistic, such as images that can be tie-pointed and mosaicked, such that they can support the development or refinement of foundational data products.
- Ensure that data are geodetically controlled as soon as possible during or after the mission so that the data are co-registered to existing foundational products.
- Consider preserving mission operations pipeline software.
- Ensure that derived products used by the team (e.g., high-level maps, mosaics, GIS layers, etc.) are delivered publicly as soon as possible during or after the mission, on the PDS Annex or some other location designated and maintained by NASA.

**Recommendation II:** NASA-funded projects, including missions and Research and Analysis projects, that obtain or create spatial data should be encouraged to deliver data in formats that are easily usable and that conform to standards agreed upon by the community. NASA and the community should work to:

- Develop a community forum for selecting and maintaining data format standards.
- Establish product formats that meet community needs for operability between products and support the integration into existing Earth-based standards.

- Encourage mission teams and research projects to incorporate data into existing Planetary Spatial Data Infrastructures that can be used by others:
    o Ensure that community standards for data formats and usability are met.
    o Ensure that sophisticated search services and visualization capabilities (such as those used by mission teams) are made available to the public.
    o Ensure that these capabilities are captured and maintained for use that continues beyond the duration of the mission or project.
    o Encourage missions and data providers to develop data user guides and other documentation and support materials for training the user.
    o Enable close cooperation between the mission teams and spatial data experts so that detailed knowledge of the instruments and their properties can be transported into the relevant PSDI

**Recommendation III:** Existing and new planetary spatial data should be easily discoverable and accessible, and data access tools must evolve with the technology. NASA and the community should work to:

- Create a clearinghouse or data portal for planetary spatial data that is easily accessible and usable (Beyer et al. 2018).
    o Ensure data are accessible via common open methods (e.g. via online Web Map Services), rather than using single tools or proprietary formats.
    o Make sure data access services use a standard Application Programming Interface (API) or can work well with a variety of interfaces.
    o Ensure this data portal technology evolves continually so that future data capabilities are represented in the data distribution services.
- Coordinate with mission teams and existing data delivery system supporters (e.g., JMARS, Solar System Treks, Small Body Mapping Tool (SBMT), etc. teams) to ensure that community data format standards are used.
    o Duplication of effort in data delivery tools should be avoided where possible, while maintaining the availability of a diverse set of tools to meet the needs of diverse user bases and use cases.
    o Data transparency and access, such as the processing steps followed for the data products, must be made available and sufficient in these tools.
- Ensure that current, high-use tools, such as web servers created by missions, entities or NASA, are maintained until a replacement is created.
- Support improvements in the quality and abundance of metadata (information about the data) for spatial data products to ensure they can be readily identified and cross-referenced in search tools in order to improve data discoverability.
- Promote a better understanding of appropriate uses for and limitations of a greatly increasing number of data products among a growing, diverse user base.

- Continue to support restoration of existing (sometimes historic) data products and metadata for high-priority targets and development of usable products from these.

**Recommendation IV:** NASA should coordinate with community representatives and groups, such as NASA Assessment Groups to ensure that foundational data products are produced and that PSDIs are developed and maintained for each planetary body in the Solar System to best enable NASA exploration and mission goals. NASA and the community should work to:
- Determine the gaps that exist in creation of control networks and PSDIs for given bodies or disciplines according to the needs of upcoming missions and exploration.
- If there are insufficient data to make necessary foundational products, identify what data are missing and how this can be addressed by future missions.
- Encourage the creation of higher-order products, such as orthoimages, mosaics, geologic maps and elevation data (topography, shape models, DEMs), to support current or upcoming exploration of planetary bodies.
- Encourage the creation of PSDIs that can support other aspects of Solar System exploration, such as human exploration, sample acquisition, analysis and return, and planetary astronomy.

**Recommendation V:** NASA and the planetary community should adopt existing standards and support the development of tools, technologies and expertise to ensure planetary spatial data are properly acquired, processed and available for effective use to the fullest extent, now and into the future. NASA and the community should work to:
- Develop a community group or special action team to review and report to NASA on needed tools, standards, technologies and expertise in spatial data analysis.
  - Look to other data user groups, especially those of Earth-based SDIs, and spatial data technologists for needed expertise and workable solutions.
- Continue to fund specialists in creating, maintaining and updating data access tools that evolve with the technology.
- Enable existing Earth-based open standards and open source software projects related to Spatial Data Infrastructures to support necessary compatibility for planetary reference systems.
- Encourage research into new ideas about spatial data manipulation and processing and data access and analytics technologies through funding of special projects on these topics.
- Provide an example of good practice to follow at an international level: help in the development of a unified system that may be applied and implemented by other space agencies (ESA, Roscosmos, JAXA, CNSA etc.) and large planetary projects (Europlanet) with useful standards and references.

- Encourage and provide an example of good practice enabling inclusion of potential increasing volumes of data from a growing number of commercial and academic payloads facilitated by increased access through commercial launch and transport.
- Ensure key areas of expertise and institutional knowledge are maintained within the fields contributing to PSDI through training and encouraging hiring and retention of spatial data experts.
    - Hold training sessions targeted for spatial data creators and users, such as those sponsored by missions at national/international conferences and at specialized data meetings and workshops.
    - Make available targeted funds to support new hires with spatial data expertise and to support recruitment fairs, scholarships and other strategies for bringing in students, postdocs and scientists. Long-term funding support at the university level might be needed (e.g., Committee (on Geodesy) 2020).

## Conclusions

The ultimate goal behind encouraging the creation of Planetary Spatial Data Infrastructures is to enable seamless discovery, access, use and fusion of spatially enabled data for all users. We would develop interfaces that exploit current technologies and evolving capabilities in pursuit of this goal, that increase scientific return, and that support NASA in its science and exploration goals. These recommendations support a broad community effort to develop tools, data products, and services for modern data storage, processing, visualization and analysis, and to deliver products that "just work" for users.

References **Archinal, B.A.,** et al. (2017). "Foundational Data Products for Europa: A Planetary Spatial Data Infrastructure Example," Fall AGU #263468; **Archinal, B.A.,** et al. (2018). "Report of the IAU Working Group on Cartographic Coordinates and Rotational Elements: 2015," Cel Mech and Dyn Ast, 130:22, DOI:10.1007/s10569-017-9805-5; **Beyer, R.**, T. Hare and J. Radebaugh (2018). The Need for a Planetary Spatial Data Clearinghouse, PSIDA #6067.pdf; **Committee** (2020). "Evolving the Geodetic Infrastructure to Meet New Scientific Needs," The National Academies Press, Washington, DC. 25579; **Duxbury, T. C.**, et al. (2002). Mars Geodesy/Cartography Working Group Recommendations on Mars Cartographic Constants and Coordinate Systems, ISPRS, 4, #521.pdf; **Gaddis, L.**, J. Laura and R. Arvidson (2018) The Role of the Planetary Data System in a Planetary Spatial Data Infrastructure, 49th LPSC, abs. #1540; **Hargitai, H.,** K. Willner and T. Hare (2019) Fundamental Frameworks in Planetary Mapping: A Review. *In*, H. Hargitai (ed.), Planetary Cartography and GIS, 978-3-319-62849-3_4; **Laura, J. R.,** et al. (2017) Towards a Planetary Spatial Data Infrastructure, *ISPRS International Journal of Geo-Information*, 6(6), 181; doi:10.3390/ijgi6060181; **Laura, J.** and Arvidson, R. E. and Gaddis, R. L. (2018a), The Relationship Between Planetary Spatial Data Infrastructure and the Planetary Data System. PSIDA #6005.pdf; **Laura, J.**, et al. (2018b) Framework for the Development of Planetary Spatial Data Infrastructures: A Europa Case Study. Earth and Space Science, 5, 486-502. 2018EA000411; **NASA** (2014). NASA Plan for Increasing Access to the Results of Scientific Research. Digital Scientific Data and Peer-Reviewed Publications, U.S.F.O of Sci and Tech Policy. NASA_Data_Plan.pdf; **Planetary Data System** (2017). Roadmap Study Report for 2017-2026 20jun17.pdf